\documentclass[preprint]{aastex}

\usepackage{xspace}

\newcommand{\msun}{\,$M_{\odot}$\xspace}
\newcommand{\msunperyr}{\,$M_{\odot}$\,yr$^{-1}$\xspace}
\newcommand{\microns}{\,$\mu$m\xspace}
\newcommand{\degrees}{$^\circ$\xspace}
\newcommand{\sqdeg}{\,deg$^2$\xspace}
\newcommand{\percent}{\,\%\xspace}
\newcommand{\kpc}{\,kpc\xspace}
\newcommand{\pc}{\,pc\xspace}

\newcommand{\Myr}{\,Myr\xspace}
\newcommand{\av}{A$_{\rm V}$\xspace}

\shorttitle{Present-day Galactic star formation rate}
\shortauthors{Robitaille and Whitney}

\begin{document}

\title{The present-day star formation rate of the Milky-Way determined from \textit{Spitzer} detected young stellar objects}

\author{Thomas P. Robitaille\altaffilmark{1,2} and Barbara A. Whitney\altaffilmark{3}}

\email{trobitaille@cfa.harvard.edu}

\altaffiltext{1}{Harvard-Smithsonian Center for Astrophysics, 60 Garden Street, Cambridge, MA, 02138, USA}
\altaffiltext{2}{Spitzer Postdoctoral Fellow}
\altaffiltext{3}{Space Science Institute, 4750 Walnut St. Suite 205, Boulder, CO 80301, USA}

\begin{abstract}
We present initial results from a population synthesis model aimed at determining the star formation rate of the Milky-Way. We find that a total star formation rate of 0.68 to 1.45\msun/yr is able to reproduce the observed number of young stellar objects in the \textit{Spitzer}/IRAC GLIMPSE survey of the Galactic plane, assuming simple prescriptions for the 3D Galactic distributions of YSOs and interstellar dust, and using model SEDs to predict the brightness and color of the synthetic YSOs at different wavelengths. This is the first Galaxy-wide measurement derived from pre-main-sequence objects themselves, rather than global observables such as the total radio continuum, H$\alpha$, or FIR flux. The value obtained is slightly lower than, but generally consistent with previously determined values. We will extend this method in the future to fit the brightness, color, and angular distribution of YSOs, and simultaneously make use of multiple surveys, to place constraints on the input assumptions, and reduce uncertainties in the star formation rate estimate. Ultimately, this will be one of the most accurate methods for determining the Galactic star formation rate, as it makes use of stars of all masses (limited only by sensitivity) rather than solely massive stars or indirect tracers of massive stars.
\end{abstract}


\maketitle

\section{Introduction}

\label{sec:introduction}

Virtually all estimates of the current star formation rates (SFRs) for galaxies in the Universe rely on stellar population synthesis models for calibration (as reviewed by \citealt{Kennicutt:98:189}). These models predict global observables, such as the total H$\alpha$ or infrared flux, that can be compared to observations \citep[e.g.][]{Calzetti:07:870,Kennicutt:09:1672}.  This is necessary because individual forming stars cannot be resolved in any galaxy other than our own.

Until recently, even estimates of the SFR of the Milky-Way have relied on global observables. Such studies generally rely on indirect tracers of massive (O- and early-B-type) stars to determine a massive SFR. This value is then extrapolated to lower masses to derive a global SFR for the Galaxy. For example, \citet{Smith:78:65} found a value of 5\msunperyr, by making use of the fact that the integrated flux density from an \ion{H}{2} region is a direct measure of the number of ionizing photons required to maintain that \ion{H}{2} region, and is therefore an indirect measure of the number of O and early-B-type stars. A more recent example includes the estimate of 4\msunperyr from \citet{Diehl:06:45} who use a measurement of the total mass of 2.8\msun of $^{26}$Al in the Galaxy from the $\gamma$-ray flux measured by INTEGRAL as a proxy for the massive star population of the Galaxy. \citet{Misiriotis:06:113} find a value of 2.7\msunperyr, by using the total 100\microns flux of the Galaxy and a conversion factor that depends on the same population synthesis model as used for other galaxies \citep{Misiriotis:04:39}. Finally, \cite{Murray:09:} use the total free-free emission in the WMAP foreground map as a probe of the massive star population, and derive a global SFR of 1.3\msunperyr. The difference between the values may in part be due to the fact that these studies do not all assume the same initial mass function (IMF). For a recent review of the tracers and determinations of the Galactic SFR, normalized to the same IMF, we refer the reader to \cite{Calzetti:09:}.

In general, any total flux measure that is related to the SFR of a galaxy (including the Milky-Way) is completely dominated by the high mass stars, since these are responsible for virtually all of the luminosity of a galaxy. Therefore, all of these methods are fundamentally limited by an extrapolation to lower masses, and higher sensitivity observations will not improve the estimates. For example, the ionizing flux for the \ion{H}{2} regions or the $\gamma$-ray flux from $^{26}$Al mentioned above are entirely dominated by the massive stars.

In this letter we propose to measure the present-day Galactic SFR by directly observing and counting pre-main sequence stars. Estimates of the SFR of individual regions within a few hundred parsecs of the Sun have recently been derived in this way \citep{Evans:09:321}, yet no such measurement exists on the scale of the Galaxy. While Galactic plane surveys with finite sensitivity preferentially favor higher mass stars, as surveys reach higher and higher sensitivities, the IMF extrapolation to lower masses will gradually account for less and less of the resulting SFR, and the accuracy of this method will be much higher than for values derived from massive star formation indicators. In fact, this method may ultimately provide a more accurate measurement than for any other galaxy, because it is a direct measure of the very young, just formed stellar population, which is unresolved in most other galaxies.

\section{Observations}

\label{sec:observations}

In this letter we make use of the \textit{Spitzer}/IRAC GLIMPSE survey of the Galactic mid-plane \citep{Benjamin:03:953},
specifically GLIMPSE I (covering $10$\degrees $\le|\ell|\le65$\degrees and $|b|\le1$\degrees) and GLIMPSE~II (which fills in the region for $|\ell|<10$\degrees, with $|b|\le1$\degrees for $|\ell|>5$\degrees , $|b|\le1.5$\degrees for 2\degrees$<|\ell|\le5$\degrees, and $|b|\le2$\degrees for $|\ell|\le2$\degrees).
The total area covered by these two surveys is 274\sqdeg.
We use the census of intrinsically red mid-infrared sources from \citet[][hereafter R08]{Robitaille:08:2413} which includes 18,949 sources from these two surveys.

The sources from R08 were selected using simple brightness and color selection criteria, namely 13.89$\ge$[4.5]$\ge$6.50, 9.52$\ge$[8.0]$\ge$ 4.01, and [4.5]-[8.0]$\ge$1. Extended sources are excluded. In total, 18,949 sources were selected, and the completeness was estimated to be at least 65.7\percent. As shown in R08, these sources consist mostly of young stellar objects (YSOs; $50-70$\percent) and asymptotic giant branch (AGB) stars ($30-50$\percent), and may include a small number of planetary nebulae (at most $2-3$\percent).
Thus, there are likely to be approximately 9,500 to 13,300 YSOs in the R08 census. Taking into account the fact that the 65.7\percent completeness is a lower limit, the actual number of YSOs that would have been detected if the survey had been complete is in the range 9,500 to 20,200.

\section{Model}

To derive a Galactic SFR using the R08 census, we construct a population synthesis model for YSOs in the Galaxy, and apply the same observational constraints as for the R08 census, varying the model SFR such that the number of `detected' synthetic YSOs matches the observed number.

The model is constructed as follows. First a 3D distribution for star formation is assumed, from which the YSO positions are randomly sampled. Each YSO is then assigned a random age and mass by assuming an IMF and a lower and upper stellar age. The number of synthetic YSOs sampled is thus directly related to the SFR for that model, which is given by the total mass of synthetic YSOs divided by the difference between the upper and lower age. Each synthetic YSO is then assigned intrinsic magnitudes at IRAC wavelengths (based on the stellar mass and age), which are scaled to its distance from the Earth. A 3D dust distribution and an extinction law are assumed in order to compute the extinction along the line of sight. Finally, the number of synthetic YSOs that fall inside the GLIMPSE I and II survey areas, satisfy the brightness and color selection criteria from R08, and are point sources, is compared to the observed number, and the SFR of the model is adjusted so that the two values match. The following paragraphs describe the assumptions made for each of these steps.  

\subsection{3D YSO distribution}

\label{sec:sfrdist}

The exact 3D distribution of YSOs in the Galaxy is uncertain, but there exist statistical estimates of where we expect stars to be forming. In particular, \cite{Boissier:99:857} used the distribution of H and H$_2$ as a function of Galactocentric radius to estimate the star formation rate relative to that at the solar radius, by assuming a Schmidt-type law relating the gas density and the star formation rate  (SFR$\propto\Sigma_{\rm gas}^{1.5}$), and taking into account the rate of passage of the gas through the spiral arms. The resulting radial profile was in good agreement with the distribution of tracers of massive stars (specifically \ion{H}{2} regions, pulsars, and supernova remnants) found by previous studies. This radial function peaks around 4 to 5\kpc. The values used were obtained from digitizing the solid line in Figure 2 of \cite{Boissier:99:857}.

We assume that the azimuthal distribution of YSOs is uniform, even though most star formation occurs in spiral arms. We make this assumption because the biggest effect of including spiral arms in the model would be to cause the longitude distribution of the sources to change on small scales, but would not change the overall number of detected synthetic sources significantly. Therefore, the azimuthally symmetric distribution we assume is still statistically representative of the overall distribution of YSOs. We assume a scale height of 80\,pc, approximately the height of the Galactic thin disk.

We take the Sun to be located at a Galactocentric radius of 8.5\kpc \citep{Kerr:86:1023}, and 27\pc above the Galactic plane \citep{Chen:01:184}.

\subsection{Stellar masses and ages}

We adopt the IMF from \citet[][equations (1) and (2)]{Kroupa:01:231}. Since the IMF is being used to implicitly extrapolate the observed mass to lower masses, it is also necessary to choose a minimum and maximum mass. The exact limits used are not important, as long as virtually all the \textit{observed} YSOs fall inside that range. The SFR derived will then be the total SFR inside that range. We choose a lower mass of 0.1\msun and an upper mass of 50\msun, since we do not expect to see any objects outside these limits -- 0.1\msun sources would be too faint, and 50\msun sources are very rare (and saturated in GLIMPSE).

Recent observations suggest that primordial circumstellar disks are present for a few Myr then disappear on relatively short timescales \citep[e.g.][]{Furlan:09:1964}.
Therefore, since we are considering YSOs with infrared excesses typical of primordial circumstellar disks (and of younger stages of evolution) we can assume that most of the R08 sources are younger than a few Myr. We choose a upper limit on the age of the synthetic YSOs of 2\,Myr, and a lower limit on the age of 1,000\,yr, as we do not expect to be able to see any sources younger than this.  It is unlikely that the overall star formation has changed significantly over the last few Myr, which is a small timescale compared to the Galactic rotation period of over 200\Myr. Therefore, we do not need to worry about time variations in the SFR, in contrast to methods based on main-sequence stars, where choosing a constant SFR is a strong assumption.

\subsection{Dust}

The 3D distribution of dust in the Galaxy is the subject of active research (e.g. \citealt{Sale:09:497}). We use the distribution from  \citet[][Equations (1), (3), and Table (2)]{Misiriotis:06:113}, which is a simple double exponential model based on modeling the far-IR dust emission, and is accurate on large scales. There exist more complex models for the distribution of dust, such as that from \cite{Drimmel:01:181}  which includes spiral arms. The effect of a more clumpy extinction distribution will be to introduce scatter in the extinction values, but the overall effect is not likely to be significant.

In recent years there have been a number of determinations of the extinction law at mid-infrared wavelengths for both the interstellar medium \citep{Lutz:99:623,Indebetouw:05:931} and for star forming regions \citep{Flaherty:07:1069}. For this work, we are modeling the large-scale extinction from the interstellar medium and therefore we use the values from \citet{Indebetouw:05:931} for the IRAC bands.

\subsection{Intrinsic magnitudes of YSOs}

\label{sec:modelsed}

The final piece of the puzzle that is needed to construct our model is a prescription to convert the masses and ages of the sources into observable magnitudes.  Unfortunately, no simple prescription exists for this, as the brightness and color depend on the amount of circumstellar dust and the properties of the central source, which in turn are far from certain for all stellar masses and ages. Of all the assumptions made in this letter, this is likely to be most crucial one.

In \citet[][hereafter R06]{Robitaille:06:256} we computed a large set of model SEDs for YSOs with a large range of evolutionary stages (from heavily embedded sources to transitional disks) and stellar masses (from 0.1\msun to 50\msun). To sample the parameters, each source was assigned a mass and age, and the stellar temperature and radius were found from evolutionary tracks. Subsequently, the disk and envelope parameters were sampled as a function of these stellar parameters. The sampling attempted to take into account observational and theoretical constraints on pre-main-sequence evolution.
It is important to note that the parameter space in R06 does not constitute an evolutionary scenario, but rather allows for different scenarios, some of which may be unrealistic. However, the general trends are nevertheless close to how we believe stars form.

For each synthetic YSO, we choose the model with the closest stellar mass and age, and sample a random viewing angle. Since we are assuming that the sources in the R08 census are younger than 2\,Myr, the R06 models with ages between 2 and 10\,Myr are not used. The IRAC magnitudes of the models we do use are dominated by the stellar luminosity and temperature, and the envelope infall rate (which determines the envelope mass and therefore the degree of obscuration of the central star). Therefore, the most important assumptions from R06 that will impact this work are the choice of stellar evolutionary tracks (which fix the stellar luminosity and temperature for a given age and mass) and the choice of the time dependence of the envelope infall rate. The latter is sampled from a range that is constant before 0.1\,Myr and decreases to zero by 1\,Myr.

The R06 models were computed in 50 circular apertures from 100 to 100,000\,AU, effectively providing radial brightness profiles at all wavelengths. Therefore, in the population synthesis model, we can find the flux for each synthetic source in an aperture corresponding to the resolution of the IRAC observations (2\arcsec). In addition,  we use this information to reject sources that would have been extended in the GLIMPSE survey, and would therefore have been excluded from the R08 census. We do this by selecting only sources where 99\percent of the flux falls inside an aperture 2\arcsec~in radius. This reduced the number of final `detected' YSOs by approximately 10\percent.

\section{Results}

\label{sec:results}

\begin{figure}[t]
\begin{center}
\includegraphics{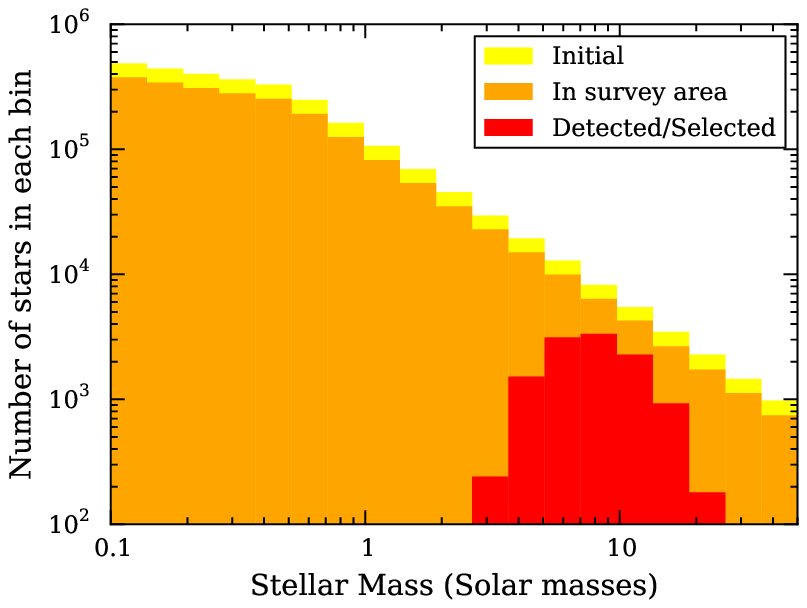}
\caption{The mass function of the synthetic YSOs for one particular run of the population synthesis model. The yellow histogram shows all of the synthetic YSOs, the orange histogram shows only the sources that fall in the survey area, and the red shows only the sources that would have been detected by \textit{Spitzer} and included in the R08 census\label{fig:imf}}
\end{center}
\end{figure}

\begin{figure}[t]
\begin{center}
\includegraphics{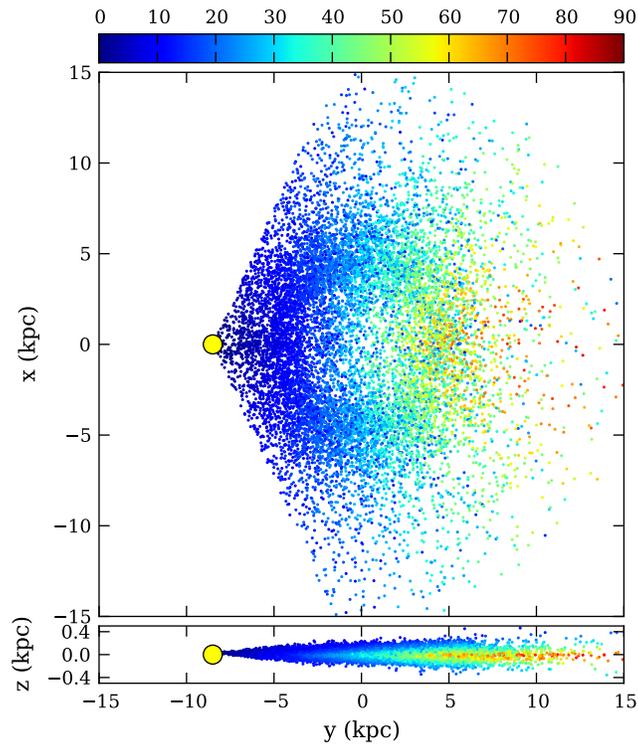}
\caption{The 3-D spatial distribution of the synthetic YSOs that would have been detected by \textit{Spitzer} and included in the R08 census, for one particular run of the population synthesis model, color-coded by the interstellar \av to the Earth (c.f. colorbar)\label{fig:spatial}}
\end{center}
\end{figure}

\begin{figure}[t]
\begin{center}
\plottwo{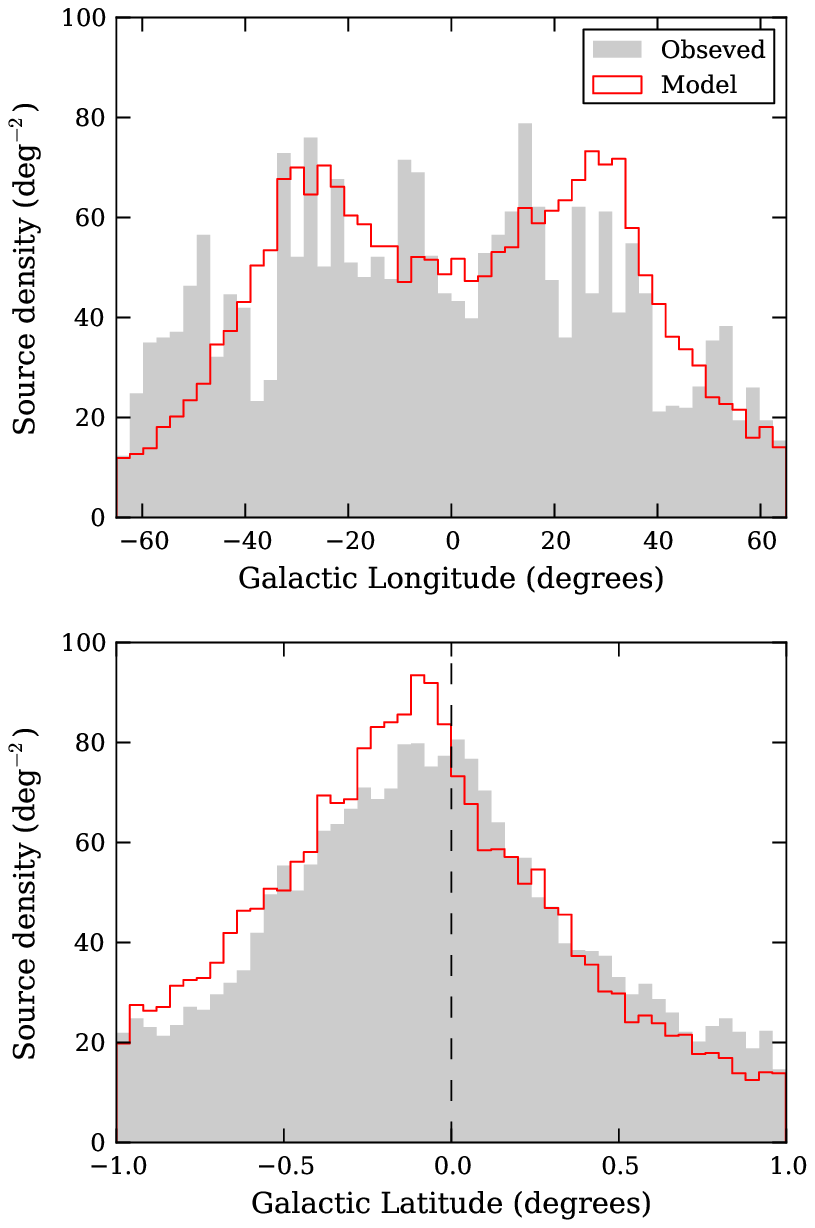}{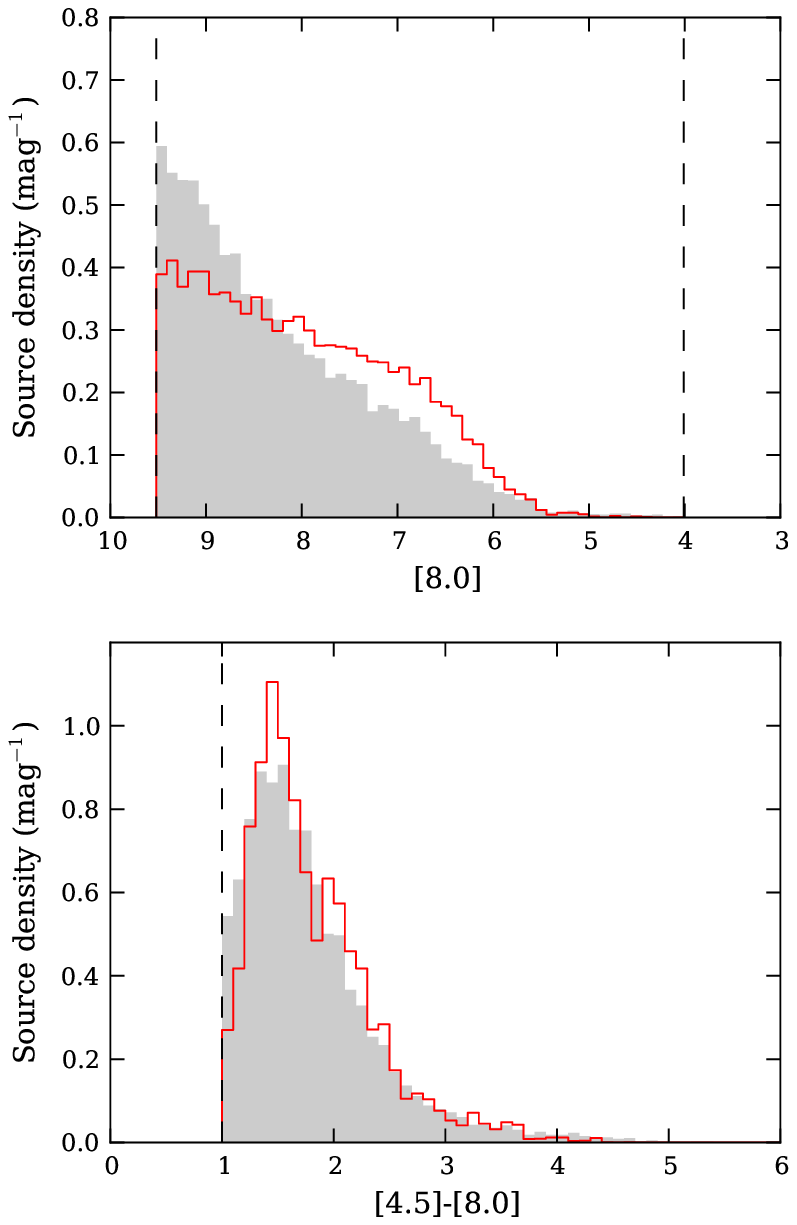}
\caption{The Galactic longitude (top left), Galactic latitude (bottom left), [8.0] magnitude (top right), and $[4.5]-[8.0]$ color (bottom right) distribution for the observed YSOs from the R08 census (gray filled histogram) and for the synthetic YSOs from one particular run of the population synthesis model (red solid line histogram). The dashed line in the Galactic latitude panel shows $b=0$\degrees to emphasize the latitude asymmetry. The dashed lines in the magnitude and color panels show the selection criteria used in R08. \label{fig:lonlat}}
\end{center}
\end{figure}

Ignoring uncertainties in the input assumption, we find that a SFR in the range 0.68 to 1.45\,\msun/yr reproduces the number of observed YSOs, where the range of values accounts for the uncertainty in the actual number of YSOs, which is in the range 9,500 to 20,200 (\S\ref{sec:observations}).
To compare the properties of the `detected' synthetic YSOs with those of the 11,649 selected candidate YSOs from R08, we computed a model with $2.73\times10^6$ synthetic YSOs -- corresponding to a SFR of 0.83\,\msun/yr -- of which 11,919 were `detected'. This model is the one shown in Figures \ref{fig:imf} to \ref{fig:lonlat}. Figure \ref{fig:imf} shows the mass distribution of all the synthetic YSOs, the subset that fall in the survey area, and the subset that would have been detected by \textit{Spitzer} and included in the R08 census. The distribution of `detected' synthetic sources is biased towards high masses, and peaks just under 10\msun. Around 50\percent of all 10\msun synthetic YSOs are detected. The drop-off at lower and higher masses is due to the lower and upper brightness cutoffs. Since approximately 2.7 million synthetic YSOs between 0.1 and 50\msun are required Galaxy-wide to explain the observed numbers, the R08 census represents less than 0.5\% of all YSOs in the Galaxy.

Figure \ref{fig:spatial} shows the 3D spatial distribution of all the synthetic YSOs that would have been included in the R08 census, color-coded by the interstellar extinction to Earth. Most sources are seen within 10\kpc but many are seen out to the far side of the Galaxy, even though the interstellar extinction is significant. The ring with a 5\kpc radius is due to the peak in the radial SFR distribution assumed (\S\ref{sec:sfrdist}), but we note that the real distribution of YSOs does not necessarily trace a ring.

Finally, Figure \ref{fig:lonlat} shows the longitude and latitude distribution, as well as the $[8.0]$ and $[4.5]-[8.0]$ distributions for all the synthetic YSOs that would have been included in the R08 census, and for all the YSO candidates in the R08 census. Even though no effort was made to fit the model distributions to the observed ones, the two are in reasonable agreement. The [8.0] magnitude distributions do diverge at the faint end, but the difference never exceeds 50\percent, and this difference may be due to remaining contaminating AGB stars. 
Both the observed and model longitude distribution drop off around longitudes of 30 to 40$^\circ$ on either side of the Galactic center. These longitudes correspond to those where the line of sight is tangent to the peak in the radial distribution of star formation (\S\ref{sec:sfrdist}). Also of interest is that the observed latitude distribution of sources is asymmetric, with $55.9\pm0.6$\percent\footnote{where the uncertainty is from Poisson statistics} of sources at negative latitudes. The model also predicts a similar asymmetry (albeit slightly larger), with $61.5\pm0.7$\percent$^1$ of sources at negative latitudes. The reason that the model presents an asymmetry is entirely due to the fact that the Sun is displaced 27\pc above the Galactic mid-plane. Therefore, it is safe to infer from this that the asymmetry in the latitude distribution of observed sources is also due to the displacement of the Sun above the Galactic mid-plane.

\section{Summary}

By developing a population synthesis model for the Milky-Way, and adjusting the star formation rate such that the number of synthetic YSOs that would have been present in the R08 census matched the actual number of observed YSOs, we obtain a star formation rate of 0.68 to 1.45\msun/yr. The uncertainties on this value take into account only the uncertainty in the actual number of observed YSOs (since these are not perfectly separable from AGB stars) and the uncertainty in the completeness. Our result is of the same order as but lower than previous estimates (\S\ref{sec:introduction}), but this can be explained in part by differences in the assumed IMF. The previous values quoted in \S\ref{sec:introduction} assume either a \cite{Salpeter:55:161} or \cite{Scalo:86:1} IMF, which over-predict the number of low-mass stars by approximately 50\percent compared to a more realistic  \cite{Kroupa:01:231} IMF. If we assume a Salpeter IMF from 0.1 to 50\msun, we obtain a SFR in the range 0.98 to 2.09\msun/yr, in better agreement with previous values. We find that the YSOs in the R08 census are dominated by sources in the range 3 to 20\msun, and that these are seen out to very large distances (10 to 15\kpc) in some cases. The longitude and latitude distribution of the synthetic YSOs are in reasonable agreement with the observed distribution, and we find that the asymmetry in the latitude distribution can be entirely explained by the fact that the Sun is not located exactly in the mid-plane.

The input assumptions for the model are simple, consisting of an axisymmetric distribution of synthetic YSOs peaking around 4 to 5\kpc, a double exponential distribution for the dust, a Kroupa IMF, a standard mid-infrared extinction law, and a prescription for the magnitudes of YSOs based on the stellar mass and age, using the models from R06. The assumption that is likely to matter the most is the latter, namely knowing what brightness and color to assign to YSOs as a function of age and mass. This is because while there are many examples of well-studied low-mass YSOs with approximate ages, the pre-main-sequence evolution of 10\msun stars, which dominate the `detected' synthetic YSOs, is poorly determined.

The model will be developed into a much more powerful diagnostic tool in the future. In particular, we plan to fit the angular, brightness, and color distributions of the sources to improve the estimate of the SFR and place constraints on other input parameters and assumptions. For example, the angular distribution of the sources will help place constraints on the model for the SFR as a function of Galactocentric radius, and the color and brightness distributions will help understand whether the assumptions for the intrinsic magnitudes of YSOs and the extinction properties are sensible, or whether they need to be modified. In addition, we will thoroughly explore the parameter space in order to understand how well constrained the input assumptions and parameters are, and what additional data could resolve degeneracies. Finally, we plan to model multiple datasets simultaneously, to reproduce not only the observable properties of sources seen in any one survey, but all available infrared or sub-mm surveys of the Galactic plane, such as the UKIDSS and Herschel surveys.

\acknowledgments

We would like to thank the anonymous referee for useful suggestions which helped improve this letter. Support for this work was provided by NASA through the Spitzer Space Telescope Fellowship (TR) and Theoretical Research (TR, BW) Programs, through a contract issued by the Jet Propulsion Laboratory, California Institute of Technology under a contract with NASA.

\clearpage

\clearpage

\clearpage

\clearpage

\end{document}